\documentclass[twocolumn,showpacs,preprintnumbers,amsmath,amssymb]{revtex4}

\usepackage{graphicx}
\usepackage{dcolumn}
\usepackage{bm}
\usepackage{epsfig}
\usepackage{longtable}

\begin{document}

\newcommand{\NB}{N\beta}
\newcommand{\EF}{E_F}

\title{Different origin of the ferromagnetic order in (Ga,Mn)As and (Ga,Mn)N}

\author{Ma\l{}gorzata Wierzbowska}  
\affiliation{%
Physics Department, Trinity College, Dublin 2, Ireland
}%
\author{Daniel S\'anchez-Portal}
\affiliation{%
Centro Mixto CSIC-UPV/EHU and
Donostia International Physics Center (DIPC), 
Paseo Manuel de Lardizabal 4, 20018 Donostia-San
Sebasti\'an, Spain
}%
\author{Stefano Sanvito}  
\affiliation{%
Physics Department, Trinity College, Dublin 2, Ireland.
}%

\date{\today}

\begin{abstract}
The mechanism for the ferromagnetic order of (Ga,Mn)As and 
(Ga,Mn)N is extensively studied over a vast range of Mn 
concentrations. 
We calculate the electronic structures of these materials using density 
functional theory in both the local spin density approximation and the 
LDA+U scheme, that we have now implemented in the code SIESTA.
For (Ga,Mn)As, the LDA+U approach leads to a hole mediated picture of the 
ferromagnetism, with an exchange constant $N\beta$=~-2.8 eV. This is smaller 
than that obtained with LSDA, which overestimates the exchange coupling 
between Mn ions and the As $p$ holes.
In contrast, the ferromagnetism in wurtzite (Ga,Mn)N is caused by the
double-exchange mechanism, since a hole of strong $d$ character is 
found at the Fermi level in both the LSDA and the LDA+U approaches. 
In this case the coupling between the Mn ions decays rapidly with the 
Mn-Mn separation. This suggests a two phases picture of the ferromagnetic
order in (Ga,Mn)N, with a robust ferromagnetic phase at large Mn concentration
coexisting with a diluted weak ferromagnetic phase.
\end{abstract}

\maketitle

\section{Introduction}

In recent years there was a rapid development 
in the growth and characterization of diluted magnetic semiconductors 
(DMS) \cite{DMSReview}. 
These materials are obtained by doping with transition metals ordinary III-V 
semiconductors \cite{Ohno35} and recently also transition metal oxides \cite{Oxides}.
The novel aspect of the DMS is the interplay between the electronic functionality
of a semiconductor with magnetic properties. For example the possibility of tailoring
the ferromagnetic Curie temperature ($T_c$) by electron gating has been already demonstrated
\cite{InAsGate}. It is then clear that the measurement of a spontaneous magnetization,
although good test for ferromagnetism, is not a direct proof of a material to be
a DMS. Therefore some other measurements such as the anomalous Hall effect
\cite{Ohno35} or X-ray magnetic circular dicrohism (XMCD) \cite{Ando} must be used to demonstrate
the interaction between the magnetic and the electronic degrees of freedom.
To date only a few materials, including (Ga,Mn)As and (In,Mn)As,
have passed convincingly this test \cite{Ando}, but unfortunately none of them 
present a $T_c$ above room temperature. This of course is a critical
requisite for future devices.

New excitement cames with the synthesis of ferromagnetic (Ga,Mn)N \cite{Sonoda1,Sonoda2,
Sonoda3,Sonoda4} with $T_c$ well above room temperature. 
High $T_c$ in this material was
somehow expected after the predictions of Dietl et al. \cite{Dietl}, who calculated the
magnetic properties of various semiconductors incorporating Mn, and concluded that
wide gap semiconductors might offer better possibility for high $T_c$. Dietl's
calculations are based on the Zener \cite{Zener} model of ferromagnetism, where 
the localized 5/2 spins of the Mn ions are antiferromagnetically coupled with the spins of the
free holes, giving rise to an effective Mn-Mn ferromagnetic interaction. 
Interestingly high $T_c$ (Ga,Mn)N does not show any evidence for anomalous hole effect, nor
for any hysteretic XMCD signal coming from the valence GaN electrons \cite{Ando2}. This seems to
suggest, that in the case of (Ga,Mn)N the agreement with the Dietl's theory is somehow 
coincidental.

Since the experimental situation is not conclusive, {\it ab-initio} methods are important for 
understanding the main features of these novel materials, and for establishing
the validity of models based on effective Hamiltonians \cite{Dietl,McDonald}.
So far a large number of density functional theory (DFT) calculations for a wide range 
of DMS have been published (see reference \cite{SSJsnm} for a review). Almost all 
the calculations to date are based on the local spin density approximation (LSDA) and 
here we list the main results for (Ga,Mn)As and (Ga,Mn)N:

\vspace{0.1in}

1) (Ga,Mn)As is a half-metal with a magnetic moment of 4$\mu_B$ per Mn in unit the 
cell \cite{SanvitoOrdejon,Mark,Leor1,Dederichs}.

\vspace{0.1in}

2) In (Ga,Mn)As the local magnetic moment at the Mn site is larger than 4$\mu_B$ and 
the Fermi level lies below the top of the majority valence band. This sustains the idea 
of a hole with spin antiferromagnetically coupled to that of the Mn. In addition an 
induced magnetic moment antiparallel to that of the Mn is found at the As sites 
neighboring the Mn ions.

\vspace{0.1in}

3) The hole in (Ga,Mn)As has a rather large $d$ component as the result of a considerable
$p$-$d$ interaction.

\vspace{0.1in}

4) For (Ga,Mn)As an estimation of the exchange constant $N\beta$ gives a value
of about -4.5~eV \cite{SanvitoOrdejon}, which is considerably larger than that 
given by most experimental determinations and the value used in model Hamiltonian
calculations \cite{Dietl, McDonald}.

\vspace{0.1in}

5) Also (Ga,Mn)N is a half-metal with a magnetization of 4$\mu_B$ per Mn in the cell
\cite{Leor}.

\vspace{0.1in}

6) The valence band of (Ga,Mn)N is not spin-splitted and the Fermi level lies in a rather
narrow impurity band \cite{Leor}.

\vspace{0.1in}

7) The magnetic impurity band in (Ga,Mn)N has a strong $d$ character and the
magnetic moment at the Mn sites is consistent with a Mn $d^4$ configuration
\cite{Mark}.

\vspace{0.1in}

Very recently we have investigated whether some of these common features 
are pathological of the use of the LSDA. In particular, since the LSDA
tends to underestimate electron localization and to overestimate the $p$-$d$ hybridization,
one may cast some doubts on its quantitative predictions. We have carried
out electronic structure calculations for both (Ga,Mn)As and (Ga,Mn)N by using the 
self-interaction corrected LDA method (LDA-SIC) \cite{Alessio1,Alessio2}. 
The main findings are that,
on the one hand the electronic structure of (Ga,Mn)As are rather similar in LSDA and 
LDA-SIC, although the second predicts a much weaker $p$-$d$ hybridization at the top of the 
valence band with consequent reduction of the valence band spin-splitting. This of course
means that the exchange constant $N\beta$ is smaller than that predicted by the LSDA.
On the other hand,  for (Ga,Mn)N LDA-SIC shows a strong orbital ordering with a convincing
evidence of a Mn $d^4$ configuration. Although this can be re-interpreted as Mn $d^5$
plus a localized $d$ hole, it is clear that no holes are left in the GaN 
valence band and an itinerant free-hole-mediated picture of ferromagnetism is 
not sustainable.

Unfortunately, due to 
its computational overheads we have not been able yet 
to investigate the details of the Mn-Mn interaction with the LDA-SIC method. 
The present paper seeks to fill this gap. We have implemented the
LDA+U scheme in the localized atomic orbital 
DFT code SIESTA \cite{SIESTA1,SIESTAcode}, 
and we have then used this novel computational capability to investigate the magnetic
properties of (Ga,Mn)As and (Ga,Mn)N. The method,
although introduces two phenomenological parameters (the Coulomb $U $and the 
exchange $J$ constants), allows us to perform large
scale calculations and therefore to investigate the Mn-Mn interaction over a broad
range of Mn concentrations.

The paper is organized as follows. In the next section we will briefly discuss our 
computational details and we will justify the values used for the LDA+U
phenomenological parameters. Then we will present our results for both (Ga,Mn)As 
and (Ga,Mn)N, and finally we will conclude. Details on the implementation of the 
LDA+U method in SIESTA are described in the appendix.
    
\section{Computational details}

All the calculations of this work are performed with the density functional 
code SIESTA \cite{SIESTA1,SIESTAcode}. 
SIESTA has been specially optimized to deal with very large systems.
It uses a very efficient localized atomic 
orbital basis set \cite{Sankey,SIESTAbasis1,SIESTAbasis2}
and norm  conserving pseudopotentials in the separate Kleinman-Bylander 
form \cite{KB}.
It is therefore ideal to simulate arrangements of hundreds and even thousands 
of atoms, hence DMS with low Mn concentrations \cite{SSJsnm} and 
related systems \cite{Zintl}.

We use conventional scalar relativistic Troulier-Martins
pseudopotentials \cite{MT} with nonlinear core corrections \cite{corr}.
The reference electronic configurations for the pseudopotentials are: 
$2s^{2}2p^{3}3d^{0}$~(N), 
$4s^{2}4p^{3}3d^{0}$~(As), $4s^{2}4p^{0}3d^{5}4f^{0}$ (Mn), with s/p/d cutoff radii:
1.14/1.14/1.14~a.u. (N), 1.9/2.18/2.5~a.u. (As) and s/p/d/f radii for Mn 
1.98/2.18/1.88/1.88~a.u.
We treat the 4$s$ and 4$p$ electrons of Ga as valence electrons and we leave
the 3$d$ in the core. Therefore the pseudopotential is constructed for
$4s^{2}4p^{1}3d^{0}$ with s/p/d cut-off radii 2.1/2.5/2.98~a.u. 
We also checked whether or not the inclusion of
$3d$ electrons in the valence changes the relevant properties.
In GaAs, although this was shown to be important for the geometry optimization and the 
high pressure phases \cite{3d-Ga}, it does not seem to be particularly relevant for the
physics at the Fermi level under normal pressure conditions. In contrast, in GaN
the 3$d$ states lie about 3~eV below the N-$2s$ states \cite{3d-exp}. LSDA sets 
erroneously their position within the N-$2s$ band and one may suspect that this will
affect somehow the physics at the Fermi level. We fix this wrong alignment by
applying $U$ corrections to the Ga $3d$ states and find that, although now the bands
have the correct position, the Fermi sphere is not modified.
Therefore we decided not to include Ga $3d$ states in the valence in order to save 
computational time and memory. Note that the same choice is presently adopted in most
of the calculations of this type \cite{Leor}.

As far as the lattice structure is concerned, all the calculations assume the experimental 
geometry for both zincblende GaAs ($a_0$=5.65 \AA) and wurtzite GaN ($a$=3.189~\AA\, 
$c$=5.185~\AA\ and $u$=0.377) \cite{Japan}. We then perform supercell simulations, where 
the supercells are constructed as integer multiple of either the primitive or in the 
case of zincblende the cubic cell.

Turning our attention to the basis set it is worth pointing out that SIESTA uses 
a flexible multiple-$\zeta$ basis set \cite{multzeta} of numerical atomic orbitals. 
Our calculations  use double-$\zeta$ for the $s$ and 
$p$-shells of any element and a triple-$\zeta$ basis set for the Mn $3d$-shell. 
Details of the relevant basis cutoff radii and their
optimization has been already given elsewhere \cite{SanvitoOrdejon}.

Although SIESTA is a very powerful and flexible package, it includes only basic 
features to tackle magnetic systems. In particular it can only use the LSDA or the 
generalized gradient approximation (GGA) for the exchange correlation potential. 
No schemes to deal with strong electron correlations are included. Since we believe that
these may play an important r\^ole in determining the magnetic properties of III-V DMS, 
we decided to implement the LDA+U scheme in SIESTA, using the functional proposed 
by Anisimov et al. \cite{AnisimovZaanen,AnisimovLongPaper}. 
Details of the implementations are discussed in the Appendix. 

Since the LDA+U method is essentially empirical, in the sense that the values
of the Coulomb and exchange constants $U$ and $J$ must be provided,
we have explored several choices of these parameters.
Our fitting criterion was to choose $U$ and $J$ in order to obtain the best energy 
position of the Mn $d$ band compared with that of available photoemission 
data \cite{spectra}. This fit fixes $U$ and $J$ respectively to 4.5~eV and 1.0~eV.
In addition to the correct positioning of the Mn $d$ band we find
that this set of parameters also reproduces accurately the band structure of (Ga,Mn)As that 
we have obtained with the LDA-SIC method \cite{Alessio1}. Finally it is worth noting that
the results are not strongly dependent on the values of $J$ (within a reasonable range) and that
we have assumed that the same set of parameters can be used also for Mn in (Ga,Mn)N.
Here we also present some results for $U=8$~eV in order to give a better explanation 
of the trends.
 
\section{Results and discussion}

A schematic Hamiltonian for describing the main interaction in III-V DMS is as follows
\begin{equation}
{\cal{H}} ={\cal{H}}_{0} + N\beta\vec{S}\cdot \vec{s}\;,
\label{HH}
\end{equation}   
where ${\cal{H}}_{0}$ is the Hamiltonian for the host semiconductor and $N\beta$ is the exchange
constant between the Mn spin $\vec{S}$ and the spin $\vec{s}$ of some type of carriers.
When the relevant carriers are holes this Hamiltonian leads to two possible scenarios 
depending on the magnitude of $N\beta/\Delta$ where $\Delta$ is the valence 
band bandwidth. 
In the case of $N\beta\ll\Delta$ the Mn-Mn coupling is mediated by free carriers and 
can be treated in term of the mean-field Zener \cite{Zener} model. 

In contrast if $N\beta\sim\Delta$ an impurity band forms at the top of the valence 
band with a strong Mn $d$ character and the effective coupling is double-exchange like 
\cite{3models}. Finally in the case $N\beta\gg\Delta$, when the hole are strongly localized at
the Mn sites, a magnetic polaron can form and the interaction is expected to be rather 
short range and described by percolation theory \cite{3models}.
In addition to this picture, if the Mn $d$ levels lie at midgap with weak coupling
with the carriers of either the conduction or valence band of the host semiconductor,
the entire use of the Hamiltonian (\ref{HH}) can be questioned and the ferromagnetism 
might be described by a Zhang-Rice polaron \cite{polaron}.
One of the aim of this work is to distinguish between these multiple options. 

\subsection{(Ga,Mn)As: Electronic Structure}

\begin{figure}[ht]
\epsfxsize=8.0cm
\centerline{\epsffile{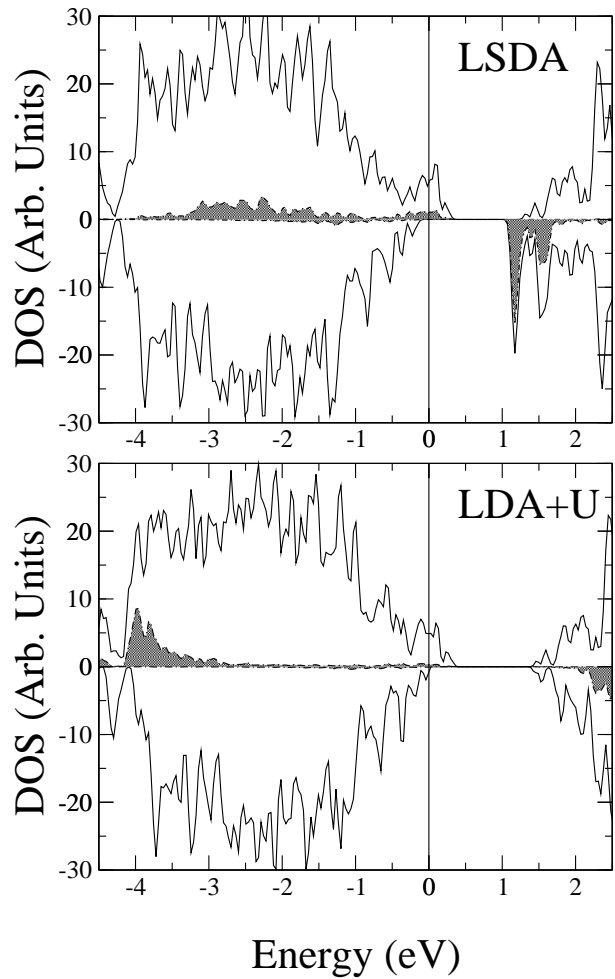}}
\caption{DOS for 3.125$\%$ (Ga,Mn)As calculated with both the LSDA (upper panel)
and the LDA+U (lower panel) schemes. The solid line represents the total DOS and the dashed
area the contribution from the Mn $d$ electrons. The values of $U$ and $J$ are respectively
4.5~eV and 1~eV. The vertical line denotes the position of the Fermi level ($E_F$=0~eV).}
\label{Fig1}
\end{figure}

In figure \ref{Fig1} we show the density of the states (DOS) for 3.125\% (Ga,Mn)As 
obtained with the LSDA and the LDA+U schemes respectively. 
Both methods result in a half-metal with a magnetic moment of 4$\mu_B$
per Mn in the cell. The difference between LSDA and LDA+U is in the position of the 
Mn-derived $d$ DOS and its contribution to the top of the majority spin valence 
band. LDA+U shifts the center on the majority Mn $d$ band to lower energies 
and now the DOS shows a strong Mn-$d$ peak at about 4~eV
below Fermi level ($E_F$). This is a feature that agrees perfectly with the photoemission 
spectra of reference \cite{spectra}. Moreover this downshift changes the contribution of the 
Mn $d$ levels to the DOS at $E_F$ which is substantial in the LSDA (18~\% of the 
total DOS at $E_F$) and small in the LDA+U (7~\%). 
These features are a consequence of the enhanced localization of the Mn $d$ shells in real space.
The magnetic moment per Mn atom calculated from the M\"ulliken population analysis 
\cite{mull} is 4.21~$\mu_B$ for LSDA and 4.71$\mu_B$ for LDA+U with $U$=4.5~eV. LDA+U
therefore gives a clear confirmation of the Mn $d^5$ configuration.

In addition there is always an induced magnetic moment at the As sites, which is
antiparallel to that of the Mn ion.
In this case the LSDA and LDA+U behaviors are somehow different. In LSDA we find a
rapid decay of the induced magnetic moment with the distance between the As and the Mn ions.
This is 0.068~$\mu_\mathrm{B}$ for the first neighbor, 0.011~$\mu_\mathrm{B}$ for the
second, 0.006~$\mu_\mathrm{B}$ for the third. In contrast LDA+U gives a larger magnetic
moment for the first neighbor (0.106~$\mu_\mathrm{B}$), but then this becomes almost constant
with the distance from the Mn ion ($\sim$0.015~$\mu_\mathrm{B}$). 

Finally it is worth noting that the LDA+U pushes the unoccupied Mn $d$ states in the minority band
to higher energies and therefore the minority conduction band 
bottom changes from $d$-like to $sp$-like.
Despite aforementioned features, it seems that taking into account strong Coulomb 
interaction in the Mn $3d$ shell, does not change much the qualitative LSDA
picture of (Ga,Mn)As near $E_F$. In order to appreciate the differences 
between the two schemes and distinguish between the different scenarios described at 
the beginning of this section we need to investigate in more details the magnetic interaction. 

\subsection{(Ga,Mn)As: Magnetic Interaction}

We first decide to evaluate the Mn-$d$/free-hole 
exchange constant $N\beta$. Before starting 
we would like to note that $N\beta$ is not a physical observable. Therefore it cannot be 
measured directly but must be inferred from some other quantity.
In particular the Hamiltonian (\ref{HH}) solved within the mean field approximation, leads
to the prediction of a linear dependence of the spin-splitting of the valence band $\Delta E_v$
upon the Mn concentration $x$
\begin{equation}
N\beta = \frac{\Delta E_v}{x \cdot \langle S \rangle }\;,
\label{NB}
\end{equation}
where $\langle S \rangle$ is the average spin of the Mn site 
(5/2 assuming a Mn $d^5$ configuration) \cite{SanvitoOrdejon}. 
This is a quantity that can be easily 
calculated within our approach. Of course the idea of extracting $N\beta$ from the 
equation (\ref{NB}) underpins the assumption that DFT and the Hamiltonian 
${\cal H}$ solved in the mean field approximation lead to the same physics. 
\begin{figure}[ht]
\epsfxsize=9.0cm
\centerline{\epsffile{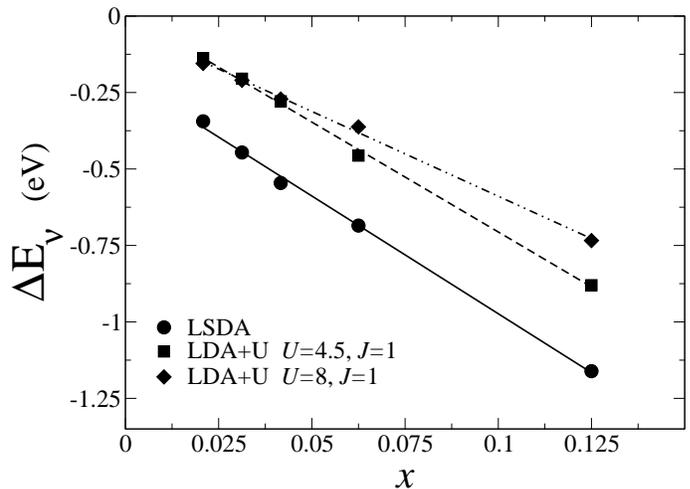}}
\caption{Valence band spin splitting $\Delta E_v$ for (Ga,Mn)As at various Mn concentrations.
Results for LSDA and LDA+U with U=4.5~eV, J=1~eV and at U=8~eV, J=1. The straight lines
represent our best fit taken over all the concentration range.}
\label{Fig2}
\end{figure}

In figure \ref{Fig2}, we present $\Delta E_v$ for $x$ equal to
0.021, 0.031, 0.042, 0.063, and 0.125 obtained by substituting one
Ga atom with Mn in supercells containing respectively 96, 64, 48, 32, and 16 atoms.
It is easy to note that the slope of the linear dependence of $\Delta E_v$ upon $x$ 
decreases when going from LSDA to LDA+U, and furthermore it becomes smaller as
$U$ gets larger. We also note that only in the case of $U$=4.5~eV the linear interpolation
presents the correct limit $\Delta E_v\rightarrow0$ for $x\rightarrow0$, while this is not found
for the LSDA or for other values of $U$. Although simple multiple scattering corrections
which reproduce the appropriate behaviour for $x\rightarrow0$ can be added to our analysis
\cite{multscatt,SanvitoOrdejon}, this is an interesting result in itself.
In fact the value of $U$=4.5~eV gives also the correct position of the Mn $d$ states, 
and produces band structure that closely agrees
with those calculated with our parameter free LDA-SIC method \cite{Alessio1}.
We therefore conclude that the LDA+U method with $U$=4.5~eV and $J$=1~eV 
accurately describes the electronic structures of (Ga,Mn)As, which in turn is also
consistent with the mean field theory \cite{Dietl}. We then extract the value
of $N\beta$ using the equation (\ref{NB}). In doing this we have observed that,
although it is possible to fit very accurately for either large
($x$=0.031, 0.042, 0.063 and 0.125) or small ($x$=0.021, 0.031, 0.042, and
0.063) Mn concentrations, a good fit over the whole concentration range is not
satisfactory. This is somehow expected since for small concentrations
multiple scattering corrections are relevant and for large the Mn ions cannot
be considered as a small perturbation to the GaAs electronic structure.
We therefore always calculate two values of $N\beta$ corresponding to the
two different fits.

In this way we find $-3.0<\NB<-2.8$~eV for LDA+U ($U=$4.5~eV), which must be put in 
relation with the values of $-3.25<\NB<-3.0$~eV and $-2.23<\NB<-2$~eV extracted respectively 
from the LSDA and from LDA+U with $U$=8~eV.

We believe that our value of $\NB\sim\:$-2.8~eV is the appropriate value to use in
the Hamiltonian (\ref{HH}). Let us point out the exact meaning of our statement.
We state that in a perfectly ordered, defect-free (Ga,Mn)As crystal, where every 
Mn ion occupies a Ga site and therefore donates a local spin 5/2 and a hole, 
the spin splitting of the valence band will be that given in figure~\ref{Fig2} 
for $U$=4.5~eV. Then, if we also suppose that $\Delta E_v$ can be described 
by the Hamiltonian ${\cal H}$ of equation (\ref{HH}), the correct ``bare'' value 
for $\NB$ is in the range $-3.0<\NB<-2.8$~eV. 
In contrast if one wants to extract 
a value of $\NB$ from experimental data 
it must be remembered 
that the actual value 
entering in the definition of the observables (for instance the band spin-splitting)
is not the ``bare'' one, but an ``effective'' value that somehow takes into account 
effects such as disorder, presence of compensating defects and so on. Therefore
it should not be surprising that these values do not agree with the ``bare'' one 
obtained here, and that the measurements of different observables give different
values of $\NB$.
Magnetotransport gives values ranging from 1.5 eV \cite{NB1} to 3.3 eV \cite{NB2},
exciton splitting gives $N\beta$=2.5~eV \cite{Szczytko} and core level photoemission 
$N\beta$=-1.2~eV \cite{NB3}.

\begin{figure}[ht]
\epsfxsize=7.0cm
\centerline{\epsffile{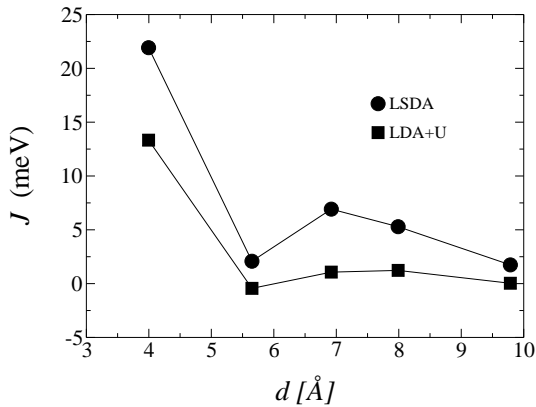}}
\caption{Exchange parameter $J$ as a function of the Mn-Mn separation $d$ for 
$x=0.063$. The results are for LSDA and LDA+U with $U$=4.5 eV and $J$=1.0.}
\label{Fig3}
\end{figure}

To further investigate the effects of strong on-site correlation on the
electronic structure of (Ga,Mn)As we have analyzed the effective Mn-Mn
interaction. As usual we consider supercells where we now
include two Mn ions per cell \cite{APLme}. We calculate the total energy 
of the cell either imposing a ferromagnetic ($E_\mathrm{FM}$) or antiferromagnetic
($E_\mathrm{AF}$) alignment of the magnetization of the Mn ions. 
Then our total energy calculations are fitted to a simple Heisenberg model, 
in which the energy can be written as
\begin{equation}
E=-\sum_{i > j}J(r_{ij})\vec{S}_i\cdot\vec{S}_j\;,
\label{heisen}
\end{equation}
where $J(r_{ij})$ is the exchange constant as a function of the Mn-Mn separation 
$r_{ij}=|\vec{R}_i-\vec{R}_j|$ and $\vec{S}_i$ is the Mn spin at site $\vec{R}_i$.
Here we assume $\vec{S}_i=\langle\vec{S}\rangle=5/2$ independently on the atomic position.
This is consistent with the magnetic moment per Mn obtained from our DFT calculations. 
Moreover we consider $J(r)$ to be rather short ranged by setting $J(r)=0$ for $r>10.0$\AA.
Our results for $x=0.063$ for both LSDA and LDA+U with $U$=4.5~eV as a function of the Mn-Mn
separation are presented in figure \ref{Fig3}.

Both LSDA and LDA+U show a decay of the magnetic coupling strength 
between two Mn ions as their separation increases. In addition we also observe some 
oscillating behaviour. This is expected when dealing with a carrier-mediated ferromagnetism, 
although our $J(r)$ cannot be fitted to an RKKY-like expression due to the lack of 
antiferromagnetic interaction (positive $J$'s).
It is interesting to note that our 
results are rather similar to those obtained previously with LSDA \cite{bruno},
which have been interpreted in term of ferromagnetic paths through As sites.
The main difference between the LSDA and the LDA+U results is a substantial 
reduction of the Mn-Mn interaction when strong correlation is included.

A rough estimation of the possible $T_c$ can be obtained by 
using the mean-field approximation \cite{DFHme}. 
This involves the sum of the exchange parameters over all the cation sites. 
From figure \ref{Fig3}
it is then clear that LDA+U predicts a $T_c$ for $\sim$6\% (Ga,Mn)As 
considerably smaller than that predicted by LSDA.

In conclusion for (Ga,Mn)As both LSDA and LDA+U give a picture of hole-mediated
ferromagnetism. However the inclusion of strong on-site repulsion at the Mn sites
reduces the $p$-$d$ hybridization at the top of the valence band with
a consequent reduction of the effective Mn-Mn interaction. 

\subsection{(Ga,Mn)N: Electronic Structure}

\begin{figure}[ht]
\epsfxsize=7.0cm
\centerline{\epsffile{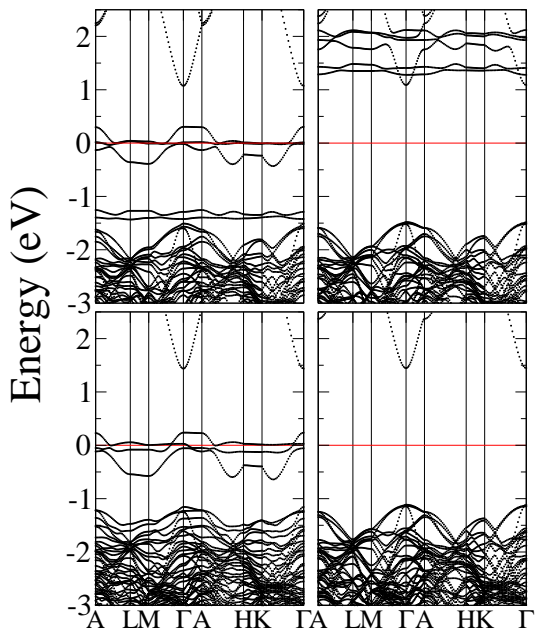}}
\caption{Bandstructure of 3.125$\%$ (Ga,Mn)N calculated with LSDA (upper panel)
and LDA+U ($U=4.5$~eV, $J=1.0$~eV) (lower panel). The picture on the left (right) is for
majority (minority) spins. The horizontal line denotes the position of the Fermi level 
($E_F$=0~eV).}
\label{Fig4}
\end{figure}

\begin{figure}[ht]
\epsfxsize=7.0cm
\centerline{\epsffile{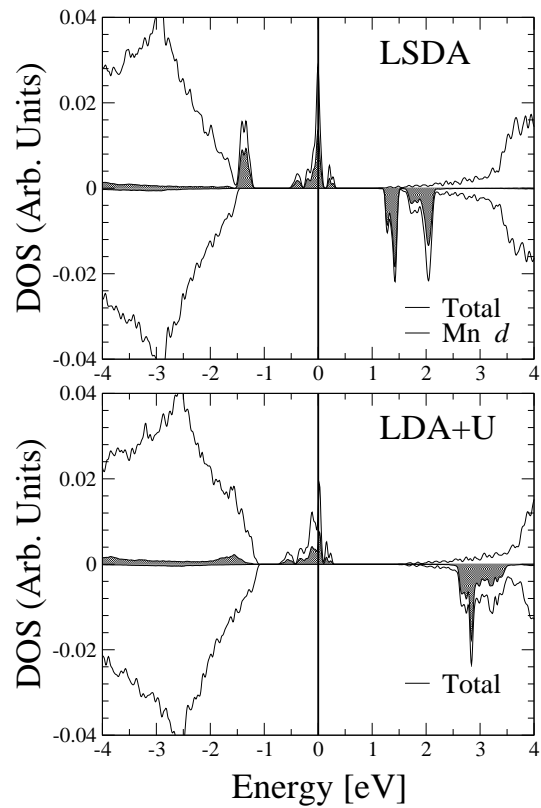}}
\caption{DOS of 3.125$\%$ (Ga,Mn)N calculated with LSDA (upper panel)
and LDA+U ($U=4.5$~eV, $J=1.0$~eV) (lower panel).
The solid line represents the total DOS and the dashed
area the contribution from the Mn $d$ electrons.
The vertical line denotes the position of the Fermi level, that we set to 0~eV.}
\label{Fig5}
\end{figure}

The bandstructure and the corresponding DOS for (Ga,Mn)N with a Mn concentration of 
3.125\% (1 Mn ion in a 64 atom GaN cell) are presented respectively in figures \ref{Fig4} and
\ref{Fig5}.
For both LSDA and LDA+U and in contrast with zincblende (Ga,Mn)As, in the wurtzite 
(Ga,Mn)N the Mn $d$-derived states appear in the middle of the GaN gap in the majority spin band. 
The Fermi level cuts 
through a triplet with mainly Mn $d$ and N $p$ character and no free holes are left in the
valence band, which in turn does not spin split. 

\begin{figure}[ht]
\epsfxsize=7.0cm
\centerline{\epsffile{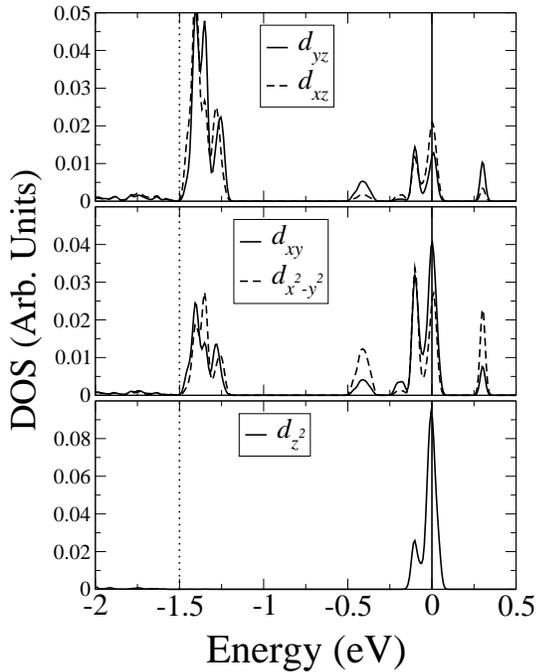}}
\caption{Projected DOS for the Mn $d$ shell of a 3.125$\%$ (Ga,Mn)N: LSDA results.
The vertical lines denote respectively the position of the Fermi level (solid line) and of the
GaN valence band top (dotted line).}
\label{Fig6}
\end{figure}

\begin{figure}[ht]
\epsfxsize=7.0cm
\centerline{\epsffile{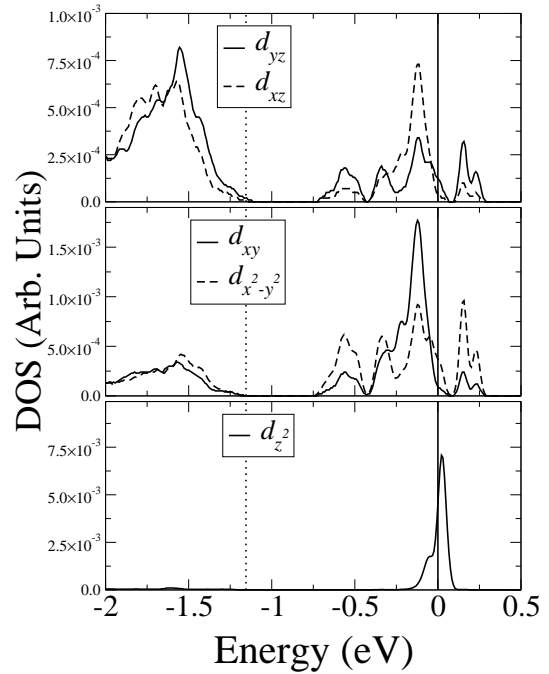}}
\caption{Projected DOS for the Mn $d$ shell of a 3.125$\%$ (Ga,Mn)N: LDA+U 
($U=4.5$~eV, $J=1.0$~eV) results. The vertical lines denote respectively the position of the 
Fermi level (solid line) and of the GaN valence band top (dotted line).}
\label{Fig7}
\end{figure}

From the pictures it is clear that also (Ga,Mn)N is a half-metal with a total magnetization 
of the unit cell of 4 $\mu_{B}$, as in the (Ga,Mn)As case. However for (Ga,Mn)N the 
M\"ulliken population gives us a magnetic moment per Mn ion of 3.74~$\mu_\mathrm{B}$
and 3.86~$\mu_\mathrm{B}$ respectively for LSDA and LDA+U. This is consistent with a Mn $d^4$ 
configuration and a Fermi surface dominated by $d$ like holes. Going into the details of the 
bandstructure we note that the triplet state at the Fermi level is made mainly from
Mn $d$ states with $xy$, $x^2-y^2$ and $z^2$ symmetry, while
the doublet is primarily made from Mn $d_{xz}$ and $d_{yz}$ states. 
The $z$-axis has been taken here along the c-axis of the wurtzite structure.
This can be easily seen from the density of states projected on the Mn $d$ shells 
(figures \ref{Fig6} and \ref{Fig7}) and it is a direct consequence of the hexagonal 
crystal field splitting. 

There are two main differences when going from LSDA to LDA+U. First with LDA+U 
the $d_{xz}$-$d_{yz}$ doublet disappears at the top of the GaN valance band, leaving
only the triplet in the bandgap, which shifts closer to the valence band top.
Secondly LDA+U enhances the orbital ordering
of the $xy$, $x^2-y^2$, $z^2$ triplet. In fact, while with LSDA this is a
mixture of all the three orbital components, with the Fermi level cutting through a 
peak with strong $d_{z^2}$ character, in LDA+U the Fermi level lies below the $d_{z^2}$ 
peak leaving a strongly localized $d_{z^2}$ hole. 
Secondly in LDA+U there is a general reduction of the Mn $d$ component of the DOS
at the Fermi level, with an increase of the N $p$ component. In fact by integrating the orbital
resolved DOS over the $xy$, $x^2-y^2$, $z^2$ triplet, we calculate that the relative contribution
to the DOS coming from the Mn $d$ shells is 55\% and 36\% respectively for LSDA and LDA+U.

These features suggest that the main effects of the Hubbard $U$ corrections are
an increase of the localization of the Mn $d$ electrons associated with a reduction of 
both the on-site $d$ hybridization and the Mn $d$-N $p$ coupling.

A similar behaviour has also been found with the LDA-SIC method \cite{Alessio1}. 
However in the LDA-SIC case the orbital ordering in much stronger producing a complete 
split of the triplet into an occupied doublet and an empty singlet (with mainly $d_{z^2}$ 
character). We have tried several values of $U$ and $J$ in order to reproduce the LDA-SIC 
result, without success. The main difference between the two calculations is that in the case of 
LDA-SIC all the orbitals, including those of Ga and N, are corrected.
The resulting band gap of GaN is almost twice as big as that obtained with the simple LSDA. 
This means that the LDA-SIC and LDA+U calculations start from an host semiconductor with a 
rather different band gap giving rise to a more pronounced orbital ordering in the 
LDA-SIC case. In support to this hypothesis it is worth reporting that novel LDA-SIC calculations \cite{Alessiopriv}, performed by subtracting the self-interaction only from the Mn $d$ orbitals, give 
very similar results than our present LDA+U.

Also for (Ga,Mn)N we investigate the presence of an induced magnetic moment at the N sites. In this
case the situation is rather different from that of (Ga,Mn)As. LSDA calculations show very
little magnetic moment at any N sites, including nearest Mn neighbors where the induced 
magnetic moment is only 0.03~$\mu_\mathrm{B}$ and parallel to that of the Mn ion. Also LDA+U
gives small induced magnetic moments (0.098~$\mu_\mathrm{B}$ for the first neighbor sites, then
smaller than 0.008~$\mu_\mathrm{B}$ for any other N), 
however those are antiparallel to that of the Mn ions.
More interestingly the occupation of the four N ions in the MnN$_4$ tetrahedron 
is rather sensitive to whether the N ion is placed along the wurtzite $c$ axis ({\it top} N) 
or in the opposite plane ({\it in-plane} N). The induced magnetic moment is
given solely by the $p_z$ orbital for the N top ion, and only by the $p_x$ and $p_y$ orbitals
for the in plane N. However, in contrast with the previous LDA-SIC results \cite{Alessio1},
in the present case the induced N magnetic moment is always antiparallel to that
of the Mn. 

\subsection{(Ga,Mn)N: Magnetic Interaction}

From the previous analysis it is clear that the bandstructure does not support a 
conventional Zener-like picture of the ferromagnetism, since no free holes are 
present in the GaN valence band. 
Therefore a careful investigation of the ferromagnetic coupling between Mn ions is needed. 
Also in this case we consider supercells where we include two Mn ions at different mutual 
positions, and we investigate the energy difference between the ferro- and antiferromagnetic 
alignment of the Mn ions.
 
\begin{figure}[ht]
\epsfxsize=7.0cm
\centerline{\epsffile{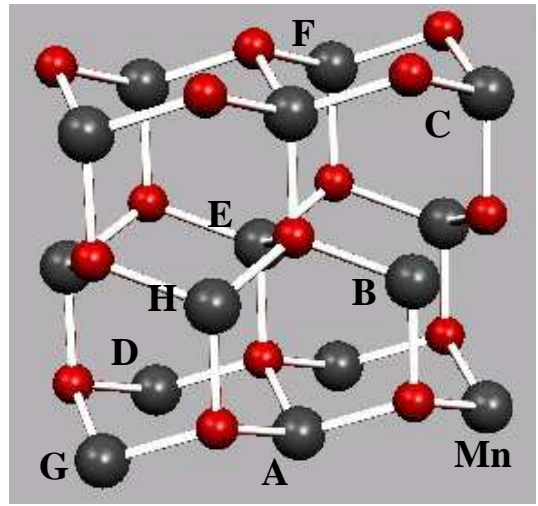}}
\caption{Position of the Mn ions in the (Ga,Mn)N supercell. The large (small) spheres
represent the possible Mn (N) sites. The first Mn ion is in the position ``Mn'' and the second occupies 
one of the sites labeled with a capital letter.}
\label{Fig8}
\end{figure}

\begin{figure}[ht]
\epsfxsize=5.5cm
\centerline{\epsffile{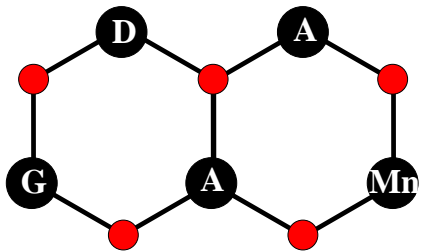}}
\caption{
Position of the Mn ions in the (Ga,Mn)N supercell: 
planar view. The large (small) circles
represent the possible Mn (N) sites. The first Mn ion 
is in the position ``Mn'' and the second occupies 
one of the sites labeled with a capital letter.}
\label{Fig9}
\end{figure}

\begin{figure}[ht]
\epsfxsize=7.0cm
\centerline{\epsffile{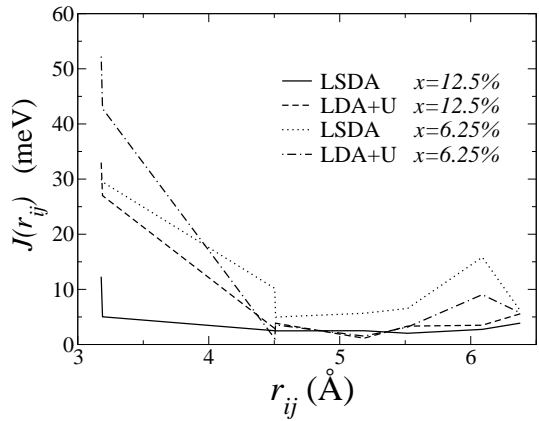}}
\caption{Exchange constants $J$ as a function of the Mn-Mn separation for $x=12.5$\% and
$x=6.25$\%, obtained with either LSDA or LDA+U. The label indicated the position of the second
Mn in the cell, according to figures \ref{Fig8} and \ref{Fig9}, and table \ref{planes}. The horizontal
line denotes $J=0$.}
\label{Fig10}
\end{figure}

We explore several possible geometrical configurations, by using both a 32 and a 64 atom unit cell.
The first Mn ion is always placed in the corner of the supercell and we allow the second to occupy 
different positions. Those possible configurations are schematically presented in figures \ref{Fig8} 
and \ref{Fig9}. Also in this case we have fit our DFT energy calculations to the Heisenberg 
Hamiltonian of equation (\ref{heisen}). In this case we use $\langle\vec{S}\rangle=2$ 
and we have set $J(r)=0$ for $r>6.4$\AA.
A summary of the results for all the Mn positions studied here is presented in table \ref{planes} 
and in figure \ref{Fig10}. 

From our data it is clear that the LDA+U gives a ferromagnetic coupling with rather different
features than that given by the LSDA. In particular when going from LSDA to LDA+U there is a large
enhancement of the $J$ for nearest Mn neighbours (case A and B), while all the other $J$'s are
either unchanged or reduced. 
A second interesting aspect is the dependence of $J(r)$ over the Mn concentration. Also in this case
LSDA and LDA+U give a rather different behaviour. In LSDA all the exchange constants
increase as the Mn concentration becomes smaller, while the LDA+U gives an increase of the
nearest neighbour constants and almost no dependence on concentration for all the other constants.

This severe dependence of the $J$'s on the Mn-Mn distance, with a strong coupling at
short Mn-Mn separation and weak long-range tails, suggests that high $T_c$ can be achieved
only at reasonably large Mn concentrations, when it is more likely to have several Mn ions occupying
nearest neighbouring positions. This leads us to speculate that the ferromagnetic state occurs in ferromagnetic clusters (with wurtzite lattice structure), presenting very large Mn concentration. 
Within the clusters the ferromagnetic state is stabilized by the strong nearest neighbour
ferromagnetic interaction, while the cluster-cluster coupling remains weak.
Therefore our LDA+U calculations support the hypothesis of two ferromagnetic phases in
(Ga,Mn)N \cite{Ando2}: a high $T_c$ phase characterized by large Mn concentrations,
and a diluted low $T_c$ phase.

\begin{table}
\begin{tabular}{cccccc}
\hline
\hline \\[-0.2cm]
  Position &  $d$ (\AA)
  & $J_{0.125}^\mathrm{LSDA}$ & $J_{0.125}^\mathrm{LDA+U}$ & $J_{0.0625}^\mathrm{LSDA}$ 
  & $J_{0.0625}^\mathrm{LDA+U}$ \\[0.1cm]
\hline \\[-0.2cm]
             A (0)    & 3.189    & 5.03  & 27.00 & 29.50 & 42.80 \\
             B (1)    & 3.179    & 12.30 & 32.99 & 30.00 & 52.23 \\
             C (2)    & 5.185    & 2.50  & 1.14  & 5.66  &  1.52 \\
             D (0)    & 4.510    & 2.48  & 3.87  & 4.98  &  3.56 \\
             E (1)    & 4.504    & 2.50  & 2.90  & 10.18 &  1.08 \\
             F (2)    & 6.088    & 2.75  & 3.50  & 15.83 &  9.05 \\
             G (0)   & 6.378    & 3.88  & 5.57  & 6.15  &  5.73 \\
             H (1)   & 5.518    & 2.04  & 3.33  & 6.55  &  3.16 \\
\hline
\hline
\end{tabular}
\caption{Summary of the results for all the different configurations studied. 
The Mn positions correspond to those of figures \ref{Fig8} and \ref{Fig9} and $d$ is
the Mn-Mn distance. 
In brackets we report the number of planes separating the two
Mn ions along the wurtzite $c$ axis. All the values of $J$ are in meV, and the
indexes
label the Mn concentration $x$.}
\label{planes}
\end{table}

\begin{figure}[ht]
\epsfxsize=7.0cm
\centerline{\epsffile{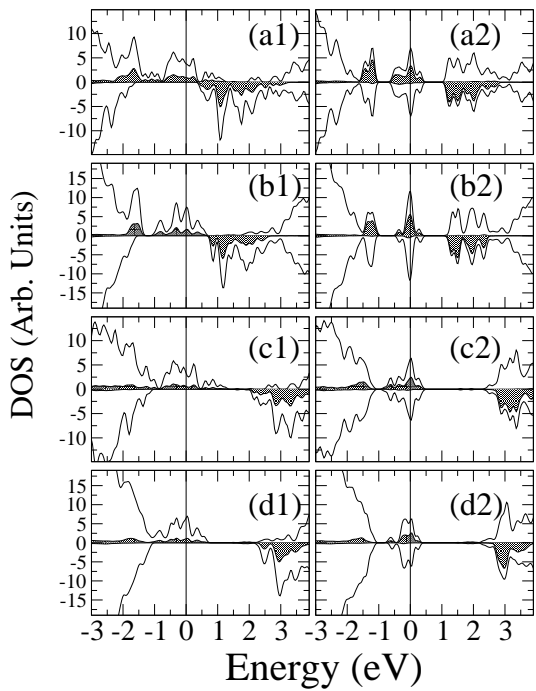}}
\caption{Density of states for various 64 and 32 atom supercells containing two Mn ions in the
position A (see figures \ref{Fig8} and \ref{Fig9}). The different figures correspond to different
Mn concentrations, to either LSDA or LDA+U and to either parallel or antiparallel alignment
of the Mn ions: (a1) $x$=12.5\% LSDA parallel, (a2) $x$=12.5\% LSDA antiparallel,
(b1) $x$=6.25\% LSDA parallel, (b2) $x$=6.25\% LSDA antiparallel,
(c1) $x$=12.5\% LDA+U parallel, (c2) $x$=12.5\% LDA+U antiparallel,
(d1) $x$=6.25\% LDA+U parallel, (d2) $x$=6.25\% LDA+U antiparallel.
The solid lines show the total DOS and the shadowed regions the contribution from the $d$ 
orbitals of one Mn ion. The horizontal lines denote the position of the Fermi level (0~eV).}
\label{Fig11}
\end{figure}

To gain more insights into the nature of the ferromagnetic coupling we have studied in 
great details the electronic structure of various supercells containing two Mn ions. 
Here we focus only on the case of nearest neighbour Mn ions, which are the ones presenting
the larger exchange constants, and in particular on the case A. In figure \ref{Fig11}
we present the DOS for both 64 and 32 atom cell containing two Mn ions in the position
A. The picture shows the results obtained with LSDA (top four panels) and LDA+U
(bottom four panels) for both the parallel (right panels) and antiparallel (left panel) 
alignments of the Mn spins. 

From the picture it is clear that the ferromagnetic state becomes more stable when
the ferromagnetic configuration presents a large spin-gap, i.e. when there is little overlap
between the two Mn derived spin sub-bands. This is the case of LDA+U for both the concentrations
studied, where we find a spin-gap of the order of $\sim$1.5~eV. Moreover in LDA+U this spin-gap 
appears to be rather insensitive to the Mn concentration (although it clearly increases when going 
from $x$=12.5\% to $x$=6.25\%), which reflects the corresponding weak dependence of $J$.

In contrast in the LSDA case there is a substantial overlap between the majority and the minority
Mn $d$ bands, in particular at large Mn concentrations. This is a direct consequence of the
broadening of the Mn $d$ states, upon increasing concentration. It is worth noting that in our 32 atom
cell each Mn ion has two Mn neighbours separated only by one N site forming a Mn-N chain, whereas
in the 64 atom cell Mn-N-Mn trimers are separated by two N sites. The large broadening obtained for
$x$=12.5\% closes almost entirely the spin gap, resulting in a rather weak ferromagnetic coupling
between the Mn. 

These findings can be understood in terms of competition between the super- and double-exchange 
mechanisms \cite{SAW}. In absence of any holes the coupling between two Mn ions
is expected to be antiferromagnetic due to the super-exchange coupling, as it was recently
verified \cite{bru_sad}. This coupling is extremely short ranged and one has to assume
that Mn ions separated by more than one N site are magnetically decoupled. Doping such a system
will generally produce a distortion of the antiferromagnetic coupling, eventually 
leading to a ferromagnetic ground state for large enough dopants concentration. 
The ``melting'' of the antiferromagnetic state is connected with the fact that the 
additional electrons (or holes) are exchange coupled with the local spin of the transition metal 
impurities. In this case the wave-function of such electrons depends on the magnetic
configuration of the transition metal ions, and in particular it will be
localized when those are aligned along different directions. The formation of a ferromagnetic
state will enhance
the delocalization of the additional electrons providing a net gain in
band energy. The final ground state is then the result of the competition between 
the energy gain due to the electron delocalization and the energy loss due to 
direct Mn-Mn super-exchange. Therefore in (Ga,Mn)N the ferromagnetic ground state is directly
connected with the presence of a hole in the Mn-derived impurity band.

However in the case of small spin-splitting of the Mn $d$ states, virtual hopping
of the holes between antiferromagnetically oriented Mn ions becomes possible
and their kinetic energy can be lowered without producing a ferromagnetic
ground state. This is why the LSDA calculations for large Mn doping present a rather
small ferromagnetic interaction between the Mn ions. In addition this enhanced hopping
between antiferromagnetically aligned Mn ions produces a reduction of the
Mn magnetic moment. For the $x$=12.5\% case LSDA shows a magnetic moment of the Mn 
$d$ shell (obtained from the M\"ulliken population) of 3.9$\mu_B$ and 3.2$\mu_B$
respectively for the ferromagnetic and antiferromagnetic configuration. This is indeed
very different from the LDA+U case, where the local magnetic moment changes little with the
mutual Mn alignment.

We have performed the same analysis over all the configurations studied with very similar
conclusions. In the other case of short Mn-Mn distance (case B) the tiny differences with
the case A are due to the different effective Mn-Mn hopping integral. This is expected since
the hopping integral depends on the specific orbitals forming the bond and ultimately on the
path connecting the two Mn ions. This gives rise to a weak
anisotropy in the magnetic coupling. Finally for Mn ions separated by more than one N site
the coupling is always rather small due to the small hopping integral. 

\section{Summary}

We have investigated extensively the mechanism for the ferromagnetic coupling in
(Ga,Mn)As and (Ga,Mn)N, by using density functional theory in both the standard LSDA and 
our newly implemented LDA+U method. 

For (Ga,Mn)As LDA+U qualitatively does not change the general picture given by LSDA.
Both methods confirm a strong $p$-$d$ hybridization leading to a spin-splitting of
the valence band of GaAs. In this case a hole mediated Zener model for the 
ferromagnetism is appropriate.
After having fixed the Coulomb and exchange constants $U$ and $J$
to values that reproduce accurately both the position of the Mn $d$ shell coming from photoemission
data \cite{spectra} and the spin-splitting of the GaAs valence band coming from the LDA-SIC 
method \cite{Alessio1}, we have estimated the Zener mean-field exchange parameter $N\beta$ to be
$\sim-2.8$~eV. We believe this is the correct value that should be used in model Hamiltonian
calculations.

Then we moved our attention to (Ga,Mn)N. In this case the addition of the on-site
$U$ corrections result in a very strong, short range, ferromagnetic coupling between 
the Mn ions. This is rather anisotropic and decays quickly with the Mn-Mn separation.
The strong ferromagnetic interaction is double-exchange-like and is associated
with the creation of a wide Mn-$d$/N-$p$ impurity band at the Fermi level. These features
sustain a picture of the ferromagnetism where a high $T_c$ ferromagnetic phase given by
regions with large Mn concentration, co-exist with a low $T_c$ ferromagnetic phase
given by small Mn concentration regions.

\begin{acknowledgments}
This work is sponsored by Enterprise Ireland under the grant EI-SC/2002/10.
Traveling is sponsored by Enterprise Ireland under the International Collaboration 
programme EI-IC/2003/47. 
DSP acknowledges support by the Basque Departamento
de Educaci\'on, the UPV/EHU (Grant. No.
9/UPV 00206.215-13639/2001), and the Spanish MCyT (Grant No.
MAT2001-09046).
\end{acknowledgments}

\begin{appendix}
\section{LDA+U implementation in the SIESTA code}

The LDA+U method combines LSDA density functional theory with the impurity Anderson 
model. The main idea is to divide the electronic states into two subsystems: localized 
(generally $d$ or $f$) and delocalized electrons (generally $s$ and $p$).
In what follows we always refer to the localized orbitals as $d$ orbitals.
Then the LDA+U philosophy consists in replacing the averaged (LSDA) Coulomb and
exchange interactions acting on the localized shells, by an orbital dependent
Hartree-Fock-like Hamiltonian \cite{AnisimovZaanen,AnisimovLongPaper}. 
The generalized LDA+U functional is defined as follows:
\begin{eqnarray}
&& E^{LDA+U}[\rho^\sigma(\vec{r}),\{n^\sigma\}]=E^{LSDA}[\rho^\sigma(\vec{r})]+E^{U}[\{n^\sigma\}]
\nonumber \\
&& -E^{dc}[\{n^\sigma\}]\;,
\end{eqnarray}
where $E^{LSDA}$ is the standard LSDA functional, $\rho^\sigma(\vec{r})$ is the charge
density for the spin $\sigma$ electrons, and $\{n^\sigma\}$ is the reduced
density matrix corresponding to the orbitals we need to correct. Finally
$E^{U}$ and $E^{dc}$ are respectively the new Hubbard-like functional and the 
double counting correction.
$E^{dc}$ is necessary to eliminate the 
averaged electron-electron interaction within
the $d$ shell, which is already included in E$_{LSDA}$. 
Following Anisimov et al. the total energy of a spin polarized system  
can be written as
\begin{eqnarray}
&& E_{LDA+U} = E_{LSDA} + \frac{1}{2} 
\sum_{mm'\sigma} 
U_{mm'}  n_{m\sigma}n_{m'-\sigma} + \nonumber \\
&&\frac{1}{2} \sum_{m\neq m'\sigma} 
(U_{mm'} -J_{mm'}) 
n_{m\sigma}n_{m'\sigma} \nonumber \\
&&- U[N^{\uparrow}(N^{\uparrow}-1)/2 
+ N^{\downarrow}(N^{\downarrow}-1)/2 + 
N^{\uparrow}N^{\downarrow}] + \nonumber \\
&& J[N^{\uparrow}(N^{\uparrow}-1)/2 + N^{\downarrow}(N^{\downarrow}-1)/2] ,
\end{eqnarray}
where $N^{\uparrow}$ and $N^{\downarrow}$ are total number of, respectively, 
spin-up and spin-down electrons 
occupying the $d$ shell,
and $n_{m\sigma}$ and $n_{m-\sigma}$ are 
the orbital occupation numbers, which are calculated self-consistently within the 
LDA+U approach. 
It is assumed in the spirit of the LDA+U approximation that the 
total occupations of the $d$
shell $N^\sigma$ are identical within LSDA and LDA+U. This assumption justifies the 
definition of the double counting term, $E^{dc}$.

The index $m$ runs over the magnetic quantum number.
The parameters $U$ and $J$ are respectively the Coulomb and the exchange 
interaction constants, that in principle can be calculated as the two-electron
matrix elements of the atomic electron-electron interaction potential, $V_{ee}$ 
\begin{eqnarray}
U_{mm'} &=& \langle mm'|V_{ee}|mm'\rangle , \\
J_{mm'} &=& \langle m'm|V_{ee}|mm'\rangle .
\end{eqnarray}
The effective LDA+U potential in then obtained by taking the 
functional derivative of the total energy $E_{LDA+U}$ with respect to the
orbital density $n_{m\sigma}({\bf r})$. This yields:
\begin{eqnarray}
&& V_{m\sigma} = V_{LSDA} - U(N-\frac{1}{2})+J(N^{\sigma}-\frac{1}{2}) +  \nonumber \\ 
&& \sum_{m' \neq m}(U_{mm'}-J_{mm'}) n_{m'\sigma} + 
\sum_{m'} U_{mm'}n_{m'-\sigma}\;.
\end{eqnarray} 
In our implementation we assume $U$ and $J$ to be independent 
from the magnetic quantum number $m$, although of course they can be different for
$d$ and $f$ shells. Hence if $U_{mm'}=U$ and $J_{mm'}=J$, the equation 
above becomes
\begin{eqnarray}
&& V_{m\sigma} = V_{LSDA} +  U\sum_{m'} (n_{m'-\sigma}-n_{0-\sigma}) + \nonumber \\ 
&& (U-J)\sum_{m' \neq m}(n_{m'\sigma}-n_{0\sigma}) + (U-J)(\frac{1}{2}-n_{0\sigma})\;,
\label{ourV}
\end{eqnarray} 
with $n_{0\sigma}$ the average orbital occupations of the correlated shell
\begin{equation}
n_{0\sigma} = \frac{1}{2l+1} N^{\sigma},
\end{equation}
and $l$ the orbital quantum number.

From the potential of equation (\ref{ourV}) one can extract an intuitive
picture of the effects of strong correlations on the one-particle energy
levels. We have 
\begin{equation}
\epsilon_{m\sigma}^{LDA+U} = \epsilon_{m\sigma}^{LSDA} + (U-J)(\frac{1}{2}-n_{m\sigma}),
\label{levels}
\end{equation}
where $n_{m\sigma}$ are the LDA+U orbital occupations. 
In the simple formula above the single particle energies of the occupied and 
unoccupied orbitals are shifted respectively by $-1/2(U-J)$ and $+1/2(U-J)$
reproducing qualitatively the correct physics of a Mott-Hubbard insulator.

We turn now our attention to the numerical implementation of this method in SIESTA.
This is quite straightforward since SIESTA uses localized atomic orbital basis set 
\cite{SIESTAcode,SIESTAbasis1}. Let us call these non-orthogonal 
basis functions $\{ \chi \}$. The two-center overlap integrals, $S_{\mu\nu}$, 
are then given by
\begin{equation} 
S_{\mu\nu} = \int \chi_{\mu}({\bf r}-{\bf R}_{1}) \chi_{\nu}({\bf r}-{\bf R}_{2}) d{\bf r},
\end{equation}
where ${\bf R}_{1}$ and ${\bf R}_{2}$ are the atomic centers, and the density
matrix in our atomic functions representation is denoted as $D_{\mu\nu}$.
The occupation number of a given atomic orbital $m$ is then defined as follows:
\begin{equation} 
n_{m\sigma} = \sum_{\mu\nu} S_{m\mu} D_{\mu\nu}^{\sigma} S_{\nu m}.
\end{equation}

The additional potential of equation (\ref{ourV}) is an operator $\hat{V}_{m\sigma}$
of the form
\begin{equation}
\hat{V}_{m\sigma}=V_{m\sigma}|m\sigma\rangle\langle m\sigma|\:,
\end{equation}
where $V_{m\sigma}$ is the scalar defined in equation (\ref{ourV}) and 
$|m\sigma\rangle\langle m\sigma|$ is the projector on the molecular state
$m$ with spin $\sigma$. Assuming $|m\sigma\rangle$ to be one of our basis 
function the matrix element of the LDA+U potential can be written as
\begin{equation}
(\hat{V}_{m\sigma})_{\mu\nu}=S_{\mu m}V_{m \sigma}S_{m\nu}\;.
\end{equation}
In this case we use a multiple-$\zeta$ basis set for the localized shell, and we
construct the LDA+U projector from one of the $\zeta$. The cut-off radius of 
this particular basis function is usually much shorter than that of the other
basis functions, and in this work we have used projectors with a cut-off radius of 2.2~Bohr.

\end{appendix}




\end{document}